\documentclass{aa}  

\usepackage{graphicx}
\usepackage{txfonts}

\newcommand{\OII}{[\mbox{O\,{\sc ii}}]}
\newcommand{\NeIII}{[\mbox{Ne\,{\sc iii}}]}
\newcommand{\OIII}{[\mbox{O\,{\sc iii}}]}
\newcommand{\NII}{[\mbox{N\,{\sc ii}}]}

\newcommand{\Ha}{H$\alpha$ }

\newcommand{\funit}{erg s$^{-1}$ cm$^{-2}$ }
\begin{document} 

   \title{The chemical enrichment of long-GRB nurseries up to $z=2$ }

\author{
S. D. Vergani\inst{1,2,3} 
\and 
J. Palmerio\inst{2} 
\and
R. Salvaterra\inst{4}
\and 
J. Japelj\inst{5}
\and 
F. Mannucci\inst{6}
\and 
D. A. Perley\inst{7}
\and
P. D'Avanzo\inst{3}
\and 
T. Kr\"uhler\inst{8}
\and
M. Puech\inst{1}
\and
S. Boissier\inst{9}
\and
S. Campana\inst{3}
\and
S. Covino\inst{3}
\and
L. K. Hunt\inst{6}
\and
P. Petitjean\inst{2}
\and
G. Tagliaferri\inst{3}
}

\institute{GEPI, Observatoire de Paris, PSL Research University, CNRS,
Univ. Paris Diderot, Sorbonne Paris Cit\'e, Place Jules Janssen, F-92195
Meudon, France; \email{susanna.vergani@obspm.fr}
\and Institut d'Astrophysique de Paris, Universit\'e Paris 6-CNRS, UMR7095, 98bis Boulevard Arago, F-75014 Paris, France
\and INAF - Osservatorio Astronomico di Brera, via E. Bianchi 46, I-23807 Merate, Italy
\and INAF - IASF Milano, via E. Bassini 15, I-20133, Milano, Italy
\and INAF - Osservatorio Astronomico di Trieste, via G. B. Tiepolo 11, I-34131 Trieste, Italy
\and INAF - Osservatorio Astrofisico di Arcetri, Largo E. Fermi 5, I-50125 Firenze, Italy
\and Dark Cosmology Centre, Niels Bohr Institute, University of Copenhagen, Juliane Maries Vej 30, DK-2100 Copenhagen, Denmark
\and Max-Planck-Institut f\"{u}r extraterrestrische Physik, Giessenbachstra\ss e, D-85748 Garching, Germany
\and Aix Marseille Univ, CNRS, Laboratoire d'Astrophysique de Marseille, F-13388, Marseille, France
             }

   \date{Received ; accepted }

 
  \abstract
   {}
   {We investigate the existence of a metallicity threshold for the production of long gamma-ray bursts (LGRBs).}
   { We used the host galaxies of the {\it Swift}/BAT6 sample of LGRBs. 
  We considered the stellar mass, star formation rate (SFR), and metallicity determined from the host galaxy photometry and spectroscopy up to $z=2$ 
and used them to compare the distribution of host galaxies to that of field galaxies in the mass-metallicity and fundamental metallicity relation plane.}
   {We find that although LGRBs also form in galaxies with relatively large stellar masses, the large majority of host galaxies have metallicities below $\log({\rm O/H})\sim8.6$. 
The extension to $z=2$ results in a good sampling of stellar masses also above \hbox{Log(M$_{*}/$M$_{\odot})\sim9.5$} and provides evidence that LGRB host galaxies do not follow the fundamental metallicity relation. 
   As shown by the comparison with dedicated numerical simulations of LGRB host galaxy population, these results are naturally explained by the existence of a mild ($\sim 0.7\;{\rm Z}_\odot$) threshold for the LGRB formation.
The present statistics does not allow us to discriminate between different shapes of the metallicity cutoff, but the relatively high metallicity threshold found in this work is somewhat in disagreement to most of the standard single-star models for LGRB progenitors. }
   {}

   \keywords{Gamma-ray burst: general -- Galaxies: abundances -- Galaxies: star formation
               }

 \maketitle
%

\section{Introduction}

It has been established that long gamma-ray bursts (LGRBs) are linked to the explosions of massive stars, both from the studies of their host galaxy formation sites \citep{Fruchter2006,Svensson2010} as well as from detections of accompanying supernova emission (GRB-SN; see \citealt{Cano2016} for a review). 
It is still not clear which conditions give rise to LGRBs or what is the relation between the progenitors of LGRBs and those of other explosions resulting from deaths of massive stars \citep[e.g.,][]{Metzger2015}.

The progenitors of nearby core-collapse supernovae can be directly identified as resolved stars in archived high-resolution images of their birth places \citep{Smartt2015}. However, LGRBs have a lower occurrence rate \citep[e.g.,][]{Berger2003,Guetta2007} and are usually observable at cosmological distances, for which their birth places cannot be resolved. Our understanding of LGRB progenitors therefore depends on linking the predictions of different stellar evolution models with the observed properties of LGRB multiwavelength emission \citep[e.g.,][]{Schulze2011,Cano2016} and their host galaxy environment (see \citealt{Perley2016a} for a review). In this work, we focus on the latter.

While metallicity is not the only factor that might affect the efficiency of the LGRB production \citep[e.g.,][]{Heuvel2013,Kelly2014,Perley2016}, it has been one of the most studied in the past as 
the metal content of the progenitor star is considered to play a major role in the formation of a LGRB explosion. 
Single-star evolution models predict that the metallicity of LGRB progenitors should be very low \citep[e.g.,][]{Hirschi2005,Yoon2005,Woosley2006}: in this way the progenitor star can expel the outer envelope (hydrogen and helium are not observed spectroscopically) without removing too much angular momentum from the rapidly rotating core. 
Higher metallicity values are allowed in the case of the models presented by \cite{Georgy2012},
also depending on the different prescriptions between the coupling of surface and core angular momentum in the star. Alternatively, the LGRB progenitors could be close interacting binaries, in which case the metallicity is a less constraining factor \citep[e.g.,][]{Fryer2007,Heuvel2007}. Strong observational constraints are clearly needed to understand which of the evolutionary channels could produce a LGRB. 

Different observational works on LGRB host galaxies in the literature have indeed revealed that their metallicities are mostly subsolar 
\citep{Modjaz2008,Levesque2010,Graham2013,Vergani2015,Kruhler2015,Perley2016,Japelj2016}.  
The evidence is corroborated by numerical simulations \citep[e.g.,][]{Nuza2007,Campisi2011,Trenti2015}. In particular, \citet{Campisi2011} studied LGRB host galaxies in the context of the mass metallicity \citep[e.g.][]{Tremonti2004} and fundamental metallicity \citep{Mannucci2010,Mannucci2011} relations of field star-forming galaxies by combining a high-resolution N-body simulation with a semi-analytic model of galaxy formation. \citet{Campisi2011} find that a very low metallicity cut is not necessary to reproduce the observed relations. 
 However, previous observational works present one or more of the following issues: (i) they are based on incomplete biased samples (e.g., \citealt{Levesque2010}); (ii) they are based on stellar masses directly determined from observations, but on metallicities inferred from the mass-metallicity relation (e.g., \citealt{Perley2016}); (iii) they use metallicities directly determined from the observations, but do not consider the stellar masses (e.g.: \citealt{Kruhler2015}); and (iv) they are based on samples limited to small redshift ranges (e.g.,\,$0<z<1$) as in \cite{Japelj2016}.

In this paper we study the metallicity of the host galaxies of the complete {\it Swift}/BAT6 sample \citep{Salvaterra2012} of LGRBs at $z < 2$, visible from the southern hemisphere. Combining the observed properties with simulations, we study their behavior in the stellar mass - metallicity relation (MZ) and fundamental metallicity relation (FMR).
After the description of the sample and new data (Section\,2), we present the results in Section\,3 and discuss them in Section\,4.

All errors are reported at 1$\sigma$ confidence unless stated otherwise. We use a standard cosmology \citep{Planck2014}: $\Omega_{\rm m} = 0.315$, $\Omega_{\Lambda} = 0.685$, and $H_{0} = 67.3$ km s$^{-1}$ Mpc$^{-1}$. The stellar masses and star formation rates (SFR) are determined using the Chabrier initial mass function \citep{Chabrier2003}.

\section{The sample}

Our sample is composed of the 27 host galaxies of the {\it Swift}/BAT6 complete sample of LGRBs at $z<2$ with declination Dec$\,<30\degr$. As the spatial distribution of GRB is isotropic, this restriction does not introduce any bias in our results. The choice to select only the LGRBs that are well observable from the southern hemisphere was due to the availability of the X-shooter spectrograph \citep{Vernet2011} at the ESO VLT (Very Large Telescope) facilities, which, thanks to its wide wavelength coverage, makes possible the detection of the emission lines necessary to  determine the SFR and metallicity of the host galaxies at $z<2$. In particular, metallicity is available for 81\% of the sample (an estimate of the metallicity was not possible for five host galaxies only).

As the original {\it Swift}/BAT6 sample is selected essentially only on the basis of the LGRB prompt $\gamma$-ray flux, and no other selection criterion is applied when gathering the galaxy sample (except the southern hemisphere visibility), our sample does not suffer of any flux bias. Indeed, no correlation has been found between the prompt $\gamma$-ray emission and host galaxy properties (see e.g.: \citealt{Levesque2010a,Japelj2016}). Furthermore, dark bursts are correctly represented in the sample (see \citealt{Melandri2012}). The restriction to the southern hemisphere at $z<2$ maintains this condition, with 26\% of LGRB of the sample being dark.

For the part of the sample at $z<1$, \cite{Vergani2015} and \cite{Japelj2016} report the tables with the objects in the sample and their properties (including stellar masses, SFR and metallicity). The restriction to the Dec\,$<30\degr$ excludes GRB\,080430 and GRB\,080319B from the sample used in this work.

 \begin{table}[!h]
 \begin{center}
      \caption[]{{\it Swift}/BAT6 sample of LGRB host galaxies at $1<z<2$ with metallicity determination, visible from the southern hemisphere. 
            }
         \label{tabsample}
           \scriptsize
         \begin{tabular}{lcccc}
\hline
\hline
Host galaxy &   redshift & Log(M${_\star}/$M${_\odot}$)   &   SFR                   & Metallicity \\
            &            &    &   [M$_{\odot}$\,yr$^{-1}$]&  12 + $\log ({\rm O}/{\rm H})$\\
 \hline
GRB080413B &1.1012     &	9.3     &	  2.1$^{+3.1}_{-1.2}$ &	 8.4$^{+0.2}_{-0.2}$ 	\\
 GRB090926B &1.2427     &	10.28   &	 26$^{+19}_{-11}$ 	 &	 8.44$^{+0.18}_{-0.20}$  \\
 GRB061007$^*$  &	 1.2623 &	9.22    &	 $5.8^{+4.8}_{-4.8} $  &     8.16 $^{+0.18}_{-0.13}$   	      \\
 GRB061121$^*$  &	 1.3160 &	10.31   &	 44.2$^{+19}_{-10}$ &	 8.5$^{+0.09}_{-0.06}$  \\
 GRB071117$^*$  &	 1.3293 &	$<10.12$&   $>2.8 $	              &          8.4$^{+0.15}_{-0.09}$  \\
 GRB100615A &1.3979     &	9.27    &	 8.6$^{+13.9}_{-4.4}$ &	 8.14$^{+0.26}_{-0.22}$ \\
 GRB070306  &	 1.4965 &	10.53   &	 101$^{+24}_{-18}$ 	  &	 8.45$^{+0.08}_{-0.08} $	\\
 GRB060306  &	 1.5597 &	10.5   	&	 17.6$^{+83.6}_{-11}$ &	 9.12$^{+0.18}_{-0.42}$	\\	      
 GRB080605  &	 1.6408 &	10.53   &	 47.0$^{+17}_{-12}$ &	 8.46$^{+0.08}_{-0.08} $	\\
 GRB080602  &	 1.8204 &	9.99    &	125.0$^{+145}_{-65}$ &	 8.56$^{+0.2}_{-0.3} $	\\
 GRB060814  &	 1.9223 &	10.82   &	 54.0$^{+89}_{-19}$ &	 8.38$^{+0.14}_{-0.28} $	\\
 \hline
 \end{tabular}
 \tablefoot{There are 4 LGRBs in the $1<z<2$ sample for which we could not determine the metallicity of their host galaxies: GRB\,050318, GRB\,050802, GRB\,060908, and GRB\,091208B. Indeed, there are no useful spectra to this purpose for the host galaxies of GRB\,091208B and GRB\,050318. For the host galaxies of GRB\,050802 and GRB\,060908 we obtained X-shooter spectroscopy (Prog. ID: 097.D-0672; PI: S.D. Vergani), but the spectra do not show sufficient emission lines to allow the metallicity determination. 

 \noindent
 $^*$: from new/unpublished X-shooter observations presented in this paper (see Table 3).
 }
 \end{center}
 \end{table}

The properties (redshift, stellar mass, SFR, and metallicity) of the $1<z<2$ part of the sample  
are reported in Table \ref{tabsample}. 
The stellar masses were taken from \cite{Perley2016}, with the exception of the host galaxies of GRB\,071117 and GRB\,080602, which are not part of the \cite{Perley2016} sample, and for which we determined the stellar masses using {\it Spitzer} observations and the same prescription as \cite{Perley2016}. 
The host of GRB\,071117 lies very close ($\sim2''$) to a red galaxy, and, therefore, the spatial resolution of the {\it Spitzer} observations allowed us to obtain only an upper limit on its infrared flux. 
We therefore also performed a spectral energy distribution fitting using the host galaxy photometry
(see Table 2) following the same prescriptions as \cite{Vergani2015}, and found log(M${_\star}/$M${_\odot}$)$\sim9.9$.

\begin{table}[!h]
\caption{Observed {\it AB} magnitudes (corrected by the Milky Way extinction) of GRB\,071117 host galaxy. }             
\label{mag}      
\centering     
\scriptsize
\begin{tabular}{l  c c c c c  }     
\hline
\hline                       
Host galaxy &  $g$ & $r$ & $i$ & $z$ & $K$ \\    
\hline     
GRB\,071117&  $24.4\pm0.1$ & $24.7\pm0.2$ & $24.8\pm0.3$ & $>24.4 $ & $22.9\pm0.2 $\\
\hline  
\end{tabular}
\tablefoot{ The $g$, $r$, $i$, $z$ magnitudes have been determined from GROND \citep{Greiner2008} observations, whereas for the $K$ value we used VLT/HAWKI observations (Prog. ID: 095.D-0560; P.I.: S.D. Vergani).
}
\end{table}

The SFR values were taken from \cite{Kruhler2015} with the exception of the host galaxies of GRB\,061007, GRB\,061121 and GRB\,071117, not included in that work. 
We obtained the VLT/X-shooter spectroscopy of these three host galaxies (ESO programs 095.D-0560 and 085.A-0795, PI: S.D. Vergani and H. Flores, respectively).
We processed the spectra using version 2.6.0 of the X-shooter data reduction pipeline \citep{Modigliani2010}, following the procedures described in \cite{Japelj2015}. 
The measured emission line fluxes are reported in Table\,\ref{tabflux}.
We determine the SFR from the \Ha fluxes 
(corrected by the extinction determined through the Balmer ratio), with the same prescriptions as \cite{Kruhler2015}. 

Following the same prescription as in \cite{Japelj2016}, we determined the metallicity of the objects in the sample with the \cite{Maiolino2008} method on the strong emission line fluxes reported in the literature \citep{Piranomonte2015,Kruhler2015} or on those measured by us; in the relevant cases, the results are consistent within errors to those already reported in the literature.

\section{The FMR and MZ relation}

   \begin{figure}[!h]
   \centering
\includegraphics[width=0.9\linewidth]{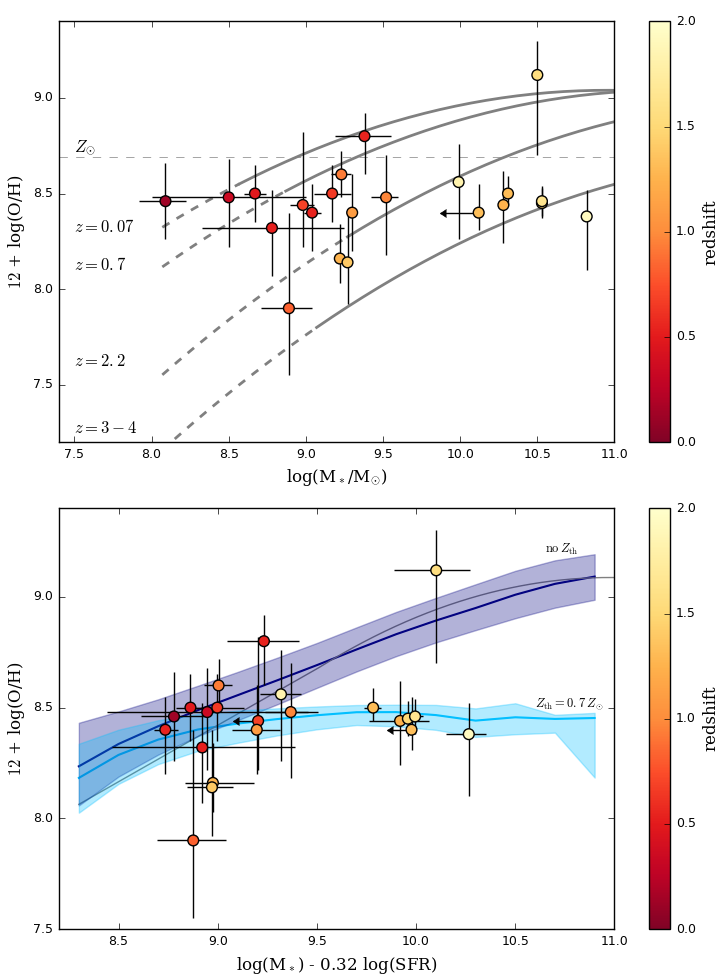}
   \caption{
   {\it Top panel:} MZ plot. 
 The dots correspond to the host galaxies of the {\it Swift}/BAT6 sample of LGRBs at $z<2$, color coded depending on their redshift as shown in the right bar. The lines correspond to the relations found for field galaxies at the redshift indicated next to each line.
 {\it Bottom panel:} The FMR plane. The dots correspond to the host galaxies of the {\it Swift}/BAT6 sample of LGRBs at $z<2$, color coded depending on their redshift as shown in the right bar. The gray line corresponds to the FMR found by \cite{Mannucci2010,Mannucci2011}. The dark blue curve and area correspond to FMR relation and of its quartiles obtained using the simulation of \cite{Campisi2011}. The cyan curve and area correspond to the best-fit model results.
   }
              \label{MZ}%
    \end{figure}

In Fig.\,\ref{MZ} we plot the host galaxies of our sample in the MZ and FMR spaces. The dearth of high metallicity galaxies is evident as well as the fact that there are more massive galaxies at the higher
redshifts ($1<z<2$) than at $z<1$.

At low stellar masses (log(M$_{*}/$M$_{\odot})<9.5$) there is some agreement with the MZ relation and FMR found for general star-forming galaxy populations (see also \citealt{Japelj2016}), whereas massive LGRB host galaxies 
are clearly shifted toward lower metallicities than predicted by the general relations. 

While the MZ relation evolves in redshift, the FMR has the advantage that it is redshift independent in the redshift range considered here, hence strengthening the statistics of our results. For the general population of star-forming galaxies with log(M$_\star)-0.32$log(SFR)$\gtrsim9.2$, the FMR is valid up to $z\sim2.2$, has been defined over SFR and stellar mass ranges encompassing those of the host galaxies in our sample, and has a smaller scatter (0.06\,dex) than the MZ relation \cite{Mannucci2010,Mannucci2011}. 
 
To verify that our results are independent of the method used to determine the metallicity, we used the \cite{Kobulnicky2004} R23 method to determine the metallicities of the 21 host galaxies for which the relevant lines to use this metallicity indicator are available. The resulting MZ plot confirms the avoidance of super-solar metallicity and the shift of high stellar mass host galaxies toward lower metallicity than those found for general star-forming galaxy populations at similar stellar masses and redshifts.

We stress that the five galaxies in the sample for which we could not determine the metallicity (GRB\,050318, GRB\,050525, GRB\,050802, GRB\,060908, and GRB\,091208B) are all faint galaxies, not hosting dark GRBs, and with stellar masses log(M$_{*}/$M$_{\odot}$)$<9.2$  (three of these galaxies have log(M$_{*}/$M$_{\odot}$)$<8.7$ ; see \citealt{Vergani2015,Perley2016}). A super-solar metallicity for a large portion of these host galaxies is therefore extremely unlikely. For two of these galaxies (GRB\,050525 and GRB\,050802) SFR limits are available (\citealt{Japelj2016}; Palmerio et al. in preparation). Under the conservative hypothesis that they follow the FMR relation, we can derive limits on their metallicities from their SFR and stellar masses of $12 + \log ({\rm O}/{\rm H})<8.1, 8.4$, respectively.

\begin{table*}[]
\centering
\scriptsize
\caption{Emission line fluxes (corrected for MW absorption) of the host galaxies of GRB\,061007, GRB061121, and GRB071117 in units 10$^{-17}$\funit. Upper limits are given at the 3$\sigma$ confidence level. 
}
\label{tabflux}
\begin{tabular}{lcccccccccc}
\hline \hline 
Host galaxy & \OII $\lambda$3726 & \OII $\lambda$3729 & \NeIII $\lambda$3869 & H$\delta$& H$\gamma $  & H$\beta $  & \OIII $\lambda$4959 &  \OIII $\lambda$5007 &H$\alpha$ & \NII $\lambda$6583  \\ 
\hline
GRB061007 & -$^{(a)}$ & $2.4\pm{0.3}$ & $<$0.7  &$<$0.7 & $<1.7$ & 1.0$\pm{0.4}$& 1.3$\pm{0.8}$ & 9.5$\pm{1.4}$ &   4.4$\pm{0.4}$& $<$2.4   \\
GRB061121 &   8.3$\pm{1.0}$ & 18.4$\pm{1.0}$ &  2.5$\pm{0.5}$ & 0.7$\pm{0.2}$ &  4.2$\pm{1.4}$  &  7.9$\pm{1.6}$  & 7.9$\pm{1.6}$ & 26.6$\pm{1.4}$ &  40.0$\pm{0.9}$& 4.5$\pm{0.8}$  \\
GRB071117 &   2.0$\pm{0.7}$ & 3.4$\pm{0.3}$ &  $<$0.4 & $<$0.8 & -$^{(b)}$ & -$^{(b)}$ &  3.0$\pm{0.6}$ & 6.6$\pm{1.0}$  & 5.6$\pm{1.0}^{(c)}$& $<$1.2  \\
\hline
\end{tabular}
\tablefoot{(a). Line strongly affected by a sky line. To determine the host galaxy properties we fixed its value to \OII $\lambda$3729$/1.5$ (low electron density case; \citealt{Osterbrock1989}). (b) Lines falling on too noisy regions to determine a significant upper limit. (c). The line is contaminated by a sky line. The flux has been determined by a Gaussian fit, using the part of the line not contaminated by the sky. 
}

\end{table*}
 
We further investigate the implications of our observational results by comparing them with the expectations of a dedicated numerical simulation of the LGRB host galaxy population presented in \cite{Campisi2009,Campisi2011}, coupling high resolution numerical simulation of dark matter with the semi-analytical models of galaxy formation described in \cite{De-Lucia2007}. Previous work \citep{De-Lucia2004} has shown that the simulated galaxy population provides a good match with the observed local galaxies properties and relations among stellar mass, gas mass and metallicity. Moreover, \cite{Campisi2011} shows that the simulations nicely reproduce the observed FMR of SDSS galaxies and its spread.
Following \cite{Campisi2011} we compute the expected number of LGRBs hosted in each simulated galaxy, assumed to be proportional to the number of short-living massive stars (i.e., star particles less than $5\times 10^7$ yr in age), applying different metallicity thresholds ($Z_{th}$) for the GRB progenitor, with probability equal to one below $Z_{th}$ and zero otherwise. 
We construct the FMR of simulated hosts in the redshift range $z=0.3-2$ and we determined the best-fit value of $Z_{th}$ by minimizing the $\chi^2$ against the BAT6 host data in the same redshift interval. The best-fit model (see Fig.\,1) is obtained for $Z_{th}=0.73^{+0.08}_{-0.07}\;{\rm Z}_{\odot}$ (1$\sigma$ errors). This is consistent with indirect results inferred from the distribution of the LGRB host stellar masses at $z<1$ \citep{Vergani2015} or of the infrared luminosities over a wider redshift range \citep{Perley2016}.

\section{Discussion and conclusions}

In this paper we considered the properties of the host galaxies of the complete {\it Swift}/BAT6 sample of LGRBs \citep{Salvaterra2012} that are visible from the southern hemisphere and at $z < 2$. We studied them with respect to the MZ and FMR relation of field star-forming galaxies. This is the first study considering at the same time the SFR, metallicity (both directly determined from the host galaxy spectroscopy), and stellar masses for a complete sample of LGRBs and on a large redshift range. Furthermore, we use LGRB host galaxy simulations to interpret our results.

Thanks to the sample extension to $z\approx2$, we could double the sample size compared to \cite{Japelj2016} and show for the first time that LGRB host galaxies do not follow the FMR. 
We find that LGRBs up to $z\approx2$ tend to explode in a population of galaxies with subsolar metallicity (Z\,$\sim0.5$-$0.8$\,Z$_{\odot}$). Our results are well reproduced by LGRB host galaxy simulations with a metallicity threshold for the LGRB production of Z$_{th}\sim0.7\;{\rm Z}_\odot$.

Although strong metallicity gradients ($>0.1-0.2$\,dex) are unlikely (on the basis of low-redshift, spatially resolved LGRB host galaxies observations; \citealt{Christensen2008,Levesque2011}; Kruhler et al. in preparation), we cannot exclude that they are at play in the couple of galaxies showing evidences of super-solar metallicities (as, e.g., in the case of GRB\,060306; see also \citealt{Niino2015}). 
The existence of some super-solar hosts may as well indicate, however, that the formation of LGRBs is also possible above the general threshold, although at much lower rate.
Applying smoother cutoffs to the metallicity, instead of the step function used here, shifts $Z_{th}$ toward lower values 
depending on the functional shape used. The present statistics does not allow us to discriminate between different cutoff shapes, therefore we do not go into further detail. We point out however that none of them succeed in reproducing the super-solar metallicity value. It should also be stressed that the GRB\,060306 metallicity is very uncertain with pretty large error bars.

The relatively high metallicity threshold found in this work is much higher than required from standard collapsar models (but see \citealt{Georgy2012}). 
Binary stars are a possible solution as progenitors, although detailed models studying the role of metallicity on the fates of binary stars are missing.
However, it is important to note that the metallicities determined using strong emission lines are not absolute values (see \citealt{Kewley2008}). In our case, they are relative to the \cite{Kewley2002} photoionization models on which the \cite{Maiolino2008} method is based. On the one hand, some works seem to indicate that 
those models may overestimate oxygen abundances by $\sim0.2$-$0.5$\,dex compared to the metallicity derived using the so-called {\it direct} $T_e$ method (see e.g., \citealt{Kennicutt2003,Yin2007}).
On the other hand, other works 
 (see e.g., \citealt{Lopez-Sanchez2012,Nicholls2012}) found that the oxygen abundances determined using temperatures derived from collisional-excited lines could be underestimated by  $\sim0.2$-$0.3$\,dex.
In principle, the simulations should be independent of these models and therefore the curves derived in this work from simulations should not be affected by this issue. 
 
 The $Z_{th}\sim0.7\;{\rm Z}_\odot$ threshold should not be considered, therefore, as an absolute value. Nonetheless, to be in agreement with the metallicities (Z$\le0.2\;{\rm Z}_\odot$) needed in most LGRB single massive star progenitor models, all the metallicities presented here should be systematically overestimated, most of them by at least $\sim0.5$\,dex.

\begin{acknowledgements}
This work is based in part on observations made with the Spitzer Space Telescope (programs 90062 and 11116), which is operated by the Jet Propulsion Laboratory, California Institute of Technology under a contract with NASA. SDV thanks M. Rodrigues, H. Flores and F. Hammer for useful discussions. JJ acknowledges financial
contribution from the grant PRIN MIUR 2012 201278X4FL 002. 
TK acknowledges support from a Sofja Kovalevskaja Award to Patricia Schady. We thanks G. Cupani for sharing his expertise on X-shooter data reduction.

\end{acknowledgements}

   \bibliographystyle{aa} 
   \bibliography{susy16} 

\end{document}